\documentclass[preprint,12pt]{elsarticle}

\usepackage{amssymb}
\usepackage{amsmath}

\usepackage{slashed} 
\usepackage[colorlinks=true, linkcolor=blue, citecolor=blue, urlcolor=blue]{hyperref}

\journal{Nuclear Physics B}

\begin{document}

\begin{frontmatter}



\title{The role of the pion mass on the QCD phase diagram in the $T-eB$ plane}


\author[inst1,inst2]{Chowdhury Aminul Islam} 
\author[inst3]{Mahammad Sabir Ali} 
\author[inst4]{Rishi Sharma} 

\address[inst1]{Institut für Theoretische Physik, Johann Wolfgang Goethe–Universität, Max-von-Laue-Str. 1, D, Frankfurt am Main, 60438, Germany}
\address[inst2]{Center for Astrophysics and Cosmology, University of Nova Gorica,\\ Vipavska 13, SI-5000
Nova Gorica, Slovenia}
\address[inst3]{School of Physical Sciences, National Institute of Science Education and Research,\\ Jatni, 752050, India}
\address[inst4]{Department of Theoretical Physics, Tata Institute of Fundamental Research,\\
Mumbai, 400005, India}            

\begin{abstract}
We investigated the role of the pion mass on the QCD phase diagram in the $T-eB$ plane using effective model treatment. Such treatments are able to capture the main features predicted by first principle calculations. We also employed the model to estimate the pion mass beyond which the inverse magnetic catalysis (IMC) effect disappears. The value is found to be independent of the strength of the magnetic field.
\end{abstract}

\begin{graphicalabstract} 
    \begin{center}
        \includegraphics[scale=0.35]{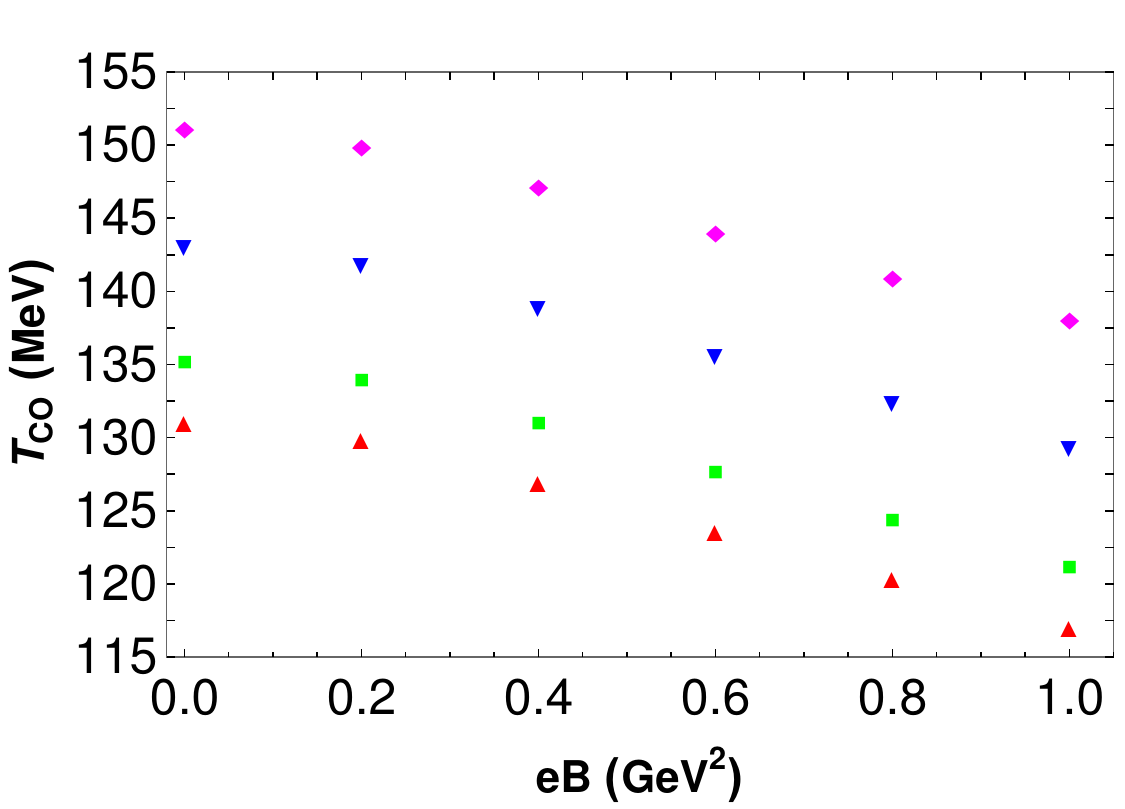}
        \includegraphics[scale=0.72]{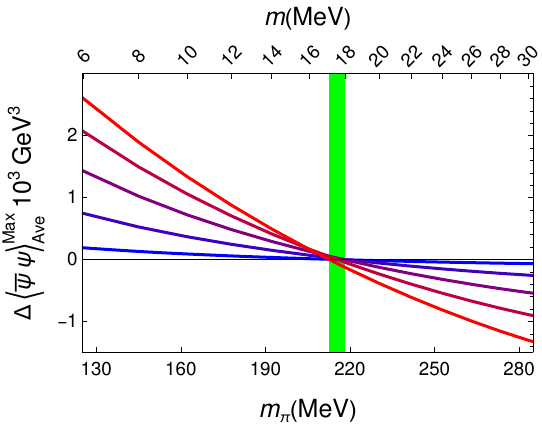}    
    \end{center}  
    The left panel captures the persistent decreasing nature of the chiral crossover temperature for increasing values of the pion mass, whereas the right panel shows a unique value of the pion mass with some uncertainty (the green band) beyond which the inverse magnetic catalysis (IMC) effect disappears in the model.
\end{graphicalabstract}

\begin{highlights}
\item An effective model treatment of the QCD has been successfully applied, for the very first time, to qualitatively capture the effect of the pion mass on the QCD phase diagram in the presence of an external magnetic field.
\item We calculated the value of the pion mass beyond which the IMC effect disappears. We have also shown that this value is independent of the strength of the magnetic field, within some uncertainty. 
\end{highlights}


\end{frontmatter}



\section{Introduction}
\label{sec:intro}

Once the discovery of the inverse magnetic catalysis (IMC) effect was made by lattice QCD~\cite{Bali:2012zg}, effective QCD models were extensively employed to reproduce it as well.~\cite{Farias:2014eca,Ferreira:2014kpa,Pagura:2016pwr}. Before the said discovery, we only knew of magnetic catalysis (MC) from the effective model calculations~\cite{Gusynin:1995nb}, which was also supported by earlier lattice QCD studies~\cite{Buividovich:2008wf}. A major group of models, namely various variations of the Nambu--Jona-Lasinio (NJL) model, utilised the running of the coupling as a function of energy to capture the IMC effect~\cite{Farias:2014eca,Ferreira:2014kpa,Pagura:2016pwr}, with some of them capturing it naturally~\cite{Pagura:2016pwr}.

Another phenomenon which accompanied the IMC effect is the reduction of the crossover temperature $(T_{\rm CO})$ with the increase of the magnetic field. This was also previously missing from our understanding in both effective model and lattice QCD studies and we always found an increasing $T_{\rm CO}$ as a function of the magnetic field~\cite{DElia:2010abb}.

Further investigations using lattice QCD revealed that the occurrence of the IMC effect depends on the values of the pion mass considered~\cite{DElia:2018xwo,Endrodi:2019zrl}. It was found that the IMC effect can disappear for sufficiently high values of the pion mass; however, the decreasing behaviour of $T_{\rm CO}$ persists with increasing pion mass. It is to be noted that a physical pion mass implies physical current quark masses; thus, heavier current quark masses correspond to heavier pion masses.

It becomes an obvious question whether the effective model treatment of QCD can capture such beyond physical point phenomena, especially after its successful reproduction of the IMC effect and the QCD phase diagram at physical point. We have found that the NJL model is capable of capturing such phenomena qualitatively~\cite{Ali:2024mnn}. In this regard, the most important feature in the model is the incorporation of the effect of a reduction in the coupling constant with increasing energy. When tested for both local and nonlocal versions, it is found that the nonlocal version captures the physics naturally and more consistently, thus leading to a successful determination of the pion mass beyond which the IMC effect disappears~\cite{Ali:2024mnn}. Here, we discuss only the results from the nonlocal treatment.

\section{Result}
\label{sec:res}

\subsection{The employed model}
\label{ssec:mod}
Here, we briefly describe the employed model. The Lagrangian for a $2$-flavor nonlocal NJL model is given as
\begin{equation}
{\cal L}_{\text{NJL}}=\bar{q}\left( i\slashed{\partial}-m\right)q+{\cal L}_{\rm sym}+{\cal L}_{\rm det},
\label{eq:mod_lag}
\end{equation}
with 
\begin{equation}
\begin{split}
{\cal L}_{\rm sym}&=G_{1}\left\{j_a(x)j_a(x)+\tilde{j}_a(x)\tilde{j}_a(x)\right\}\,\,{\rm and}\\
{\cal L}_{\rm det}&=G_{2}\left\{j_a(x)j_a(x)-\tilde{j}_a(x)\tilde{j}_a(x)\right\},
\end{split}
\label{eq:lag_diff_terms}
\end{equation}
where, $q$ is the quark doublet with light quarks $u$ and $d$. Both these quarks have equal mass $m$.  A general description of nonlocal currents in the interaction terms, ${\cal L}_{\rm sym}$ and ${\cal L}_{\rm det}$,  is 
\begin{eqnarray}
j_{a}(x)/\tilde{j}_a(x)=\int d^4z\ {\cal H}(z)\bar{q}\left(x+\frac{z}{2}\right)\Gamma_{a}/\tilde{\Gamma}_a q(x-\frac{z}{2}),
\label{eq:cur_nonlocal}
\end{eqnarray}
where $\Gamma_a/\tilde{\Gamma}_a$ contains the appropriate combination of the Pauli matrices and ${\cal H}(z)$ is the nonlocal form factor~\cite{Ali:2024mnn}. Replacing it with a delta function reduces the Lagrangian in Eq.~\eqref{eq:mod_lag} into its local version.

\subsection{Disappearance of the IMC effect}
\label{ssec:dis_imc}
\begin{figure}[h!]
\centering
\includegraphics[scale=0.6]{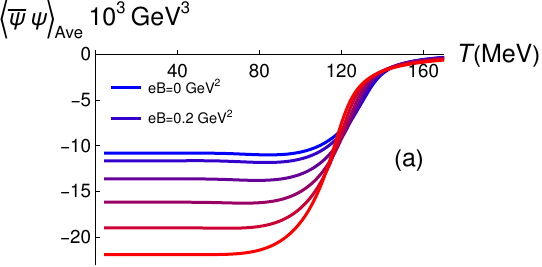}
\includegraphics[scale=0.6]{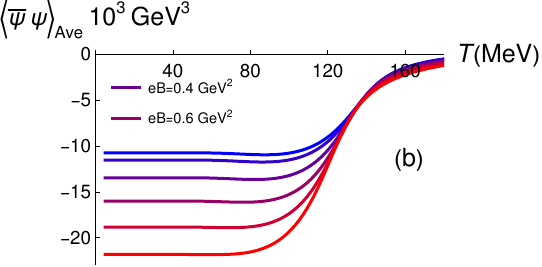}
\includegraphics[scale=0.6]{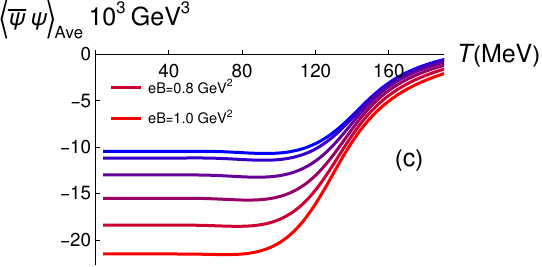}
\includegraphics[scale=0.6]{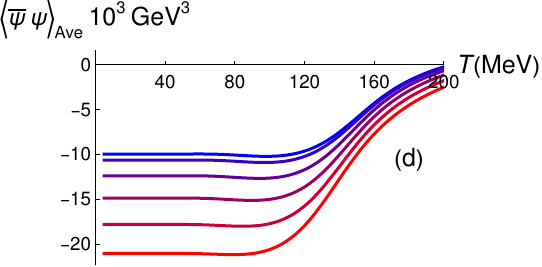}
\caption{Condensate averages for different values of the magnetic field as a function of temperature plotted for different values of the pion mass.}
\label{fig:cond_avg}
\end{figure}
In Fig.~\ref{fig:cond_avg}, the condensate averages are plotted for different values of the pion mass. The figures (a), (b), (c) and (d) correspond to pion mass values of $135$, $220$, $340$ and $440$ MeV, respectively. It is evident that the IMC effect diminishes and disappears for a higher pion mass value.

\subsection{The phase diagram}
\label{ssec:pd}
\begin{figure}[h!]
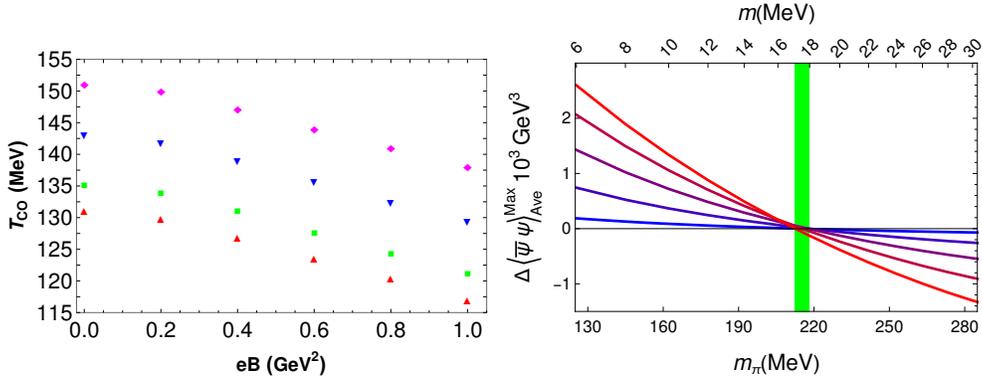

\centering
\includegraphics[scale=0.34]{PhaseDiagram.pdf}
\includegraphics[scale=0.7]{CriticalMq.pdf}
\caption{Left panel: QCD phase diagram for increasing value of the pion mass. Right panel: The specific value of the pion mass beyond which the IMC effect disappears.}
\label{fig:pd_n_crt_mpi}
\end{figure}
The QCD phase diagram is shown in the left panel of Fig.~\ref{fig:pd_n_crt_mpi}. For all the values of the pion mass tested here, the decreasing nature of the $T_{\rm CO}$ sustains.

\subsection{A distinct value of the pion mass}
\label{ssec:dis_pion}
Using the model, we estimate the pion mass or the quark mass beyond which the IMC effect disappears, as shown by the green band in the right panel of Fig.~\ref{fig:pd_n_crt_mpi}.









\section{Conclusion}
\label{sec:con}
It has been successfully demonstrated that an effective model treatment of QCD can capture the beyond physical point physics, at least qualitatively. In this case, we find, within the premises of the model, that the IMC effect disappears, whereas the decreasing trend of the crossover temperature with increasing pion mass persists, in agreement with lattice QCD results.

We further utilise the simple working mechanism and cost-effective implementation of the models to calculate the pion mass beyond which the IMC effect disappears, across different values of the magnetic field, unlike the single value tested in lattice QCD so far~\cite{Endrodi:2019zrl}. We find it to be a specific value of $215\,{\rm MeV}$ with a small spread, which is much lower than what is predicted by lattice QCD, $497(4)\,{\rm MeV}$ at $eB=0.6\,{\rm GeV^2}$. Such a magnetic-field-independent specific value of the pion mass is yet to be confirmed by lattice QCD. 

Finally, one wonders how such a simple-minded effective model treatment can capture complex phenomena predicted by first-principles QCD calculations. It then indulges one in thinking that the underlying principal mechanisms responsible for such phenomena may not be too complex!

\end{document}